\documentclass{article}
\usepackage{spconf,amsmath,graphicx}
\usepackage{amsthm}
\usepackage{url}
\usepackage{graphicx}
\usepackage{caption}
\usepackage{subcaption}
\usepackage{color,soul}
\usepackage[utf8]{inputenc}
\usepackage{multirow}
\newtheorem{theorem}{Theorem}

\newtheorem{prop}{Proposition}

\usepackage{amsfonts}
\usepackage[english]{babel}

\title{Adaptive Group Testing with Mismatched Models}
\name{Mingzhou Fan$^\star$, Byung-Jun Yoon$^{\star \dagger}$, Francis J. Alexander$^\dagger$, Edward R. Dougherty$^\star$, Xiaoning Qian$^{\star \dagger}$\thanks{The work is supported in part by the U.S. Department of Energy, Office of Science, Office of Advanced Scientific Computing Research, Mathematical Multifaceted Integrated Capability Centers program under Award Q1 DE-SC0019303, and the National Science Foundation (NSF) Awards 1553281, 1812641, and 2119103.}}
\address{$^{\star}$ Department of Electrical \& Computer Engineering, Texas A\&M University, College Station, TX\\
      $^{\dagger}$ Computational Science Initiative, Brookhaven National Laboratory, Upton, NY}
\begin{document}
%
\maketitle
\begin{abstract}
Accurate detection of infected individuals is one of the critical steps in stopping any pandemic. 
When the underlying infection rate of the disease is low, testing people in groups, instead of testing each individual in the population, can be more efficient. 
In this work, we consider noisy adaptive group testing design with specific test sensitivity and specificity that select the optimal group given previous test results based on pre-selected utility function.
As in prior studies on group testing, we model this problem as a sequential Bayesian Optimal Experimental Design (BOED) to adaptively design the groups for each test. 
We analyze the required number of group tests when using the updated posterior on the infection status and the corresponding Mutual Information (MI) as our utility function for selecting new groups. 
More importantly, we study how the potential bias on the ground-truth noise of group tests may affect the group testing sample complexity.

\end{abstract}
\section{Introduction}
\label{sec:intro}

Originally proposed for blood testing in the World War II~\cite{Doffman1943GT}, group testing has been a powerful tool for detecting infected individuals in a large population, for example by polymerase chain reaction (PCR) tests for COVID-19~\cite{Gollier2020COVID}. By mixing up the test samples (e.g. saliva or blood) of individuals in a group, the tester can determine whether there exists any infected individual in the group if no mistakes are made.

Current studies on group testing can be roughly divided into two categories -- \textit{non-adaptive}~\cite{Han2017GT, Liu2011GT, Du1993GT, Aita2021GT, Sobel1975GT} and \textit{adaptive} methods \cite{Baldassini2013GT, Han2017GT, Google2020GT, Sakata2021GT, Bai2019GT, McMahan2017GT, HughesOliver1994GT, Scarlett2019GT} -- based on  whether the group to test is decided before the tests or adaptively given test results during the whole sequential procedure. 

Non-adaptive group testing is solved in a two-stage fashion: designing testing groups and recovering the infection status based on the testing results. The very first paper about group testing \cite{Doffman1943GT} is in a non-adaptive way and derived the optimal group selection in a noiseless setup. Recovering for large population is considered in~\cite{Sobel1975GT}. 
A noisy setup was considered and the optimality of group testing was proved in~\cite{Liu2011GT}. 
A review of non-adaptive group testing and its applications can be found in~\cite{Du1993GT}. 

Adaptive group testing updates the Bayesian models of infection status based on the previous testing results and designs a new group to test at each iteration. Works on adaptive group testing have examined its empirical performance in different set-ups. 
Bayesian regression was used in~\cite{McMahan2017GT} to model the group testing data with adaptive group tests based on the updated posterior. 
Model inference by lattice-based classification models~\cite{Tatsuoka2021GT} or sum-observation~\cite{Han2017GT} has been explored. In~\cite{Sakata2021GT}, Loopy Belief Propagation (LBP)~\cite{Murphy1999LBP} and other approximation strategies~\cite{Bai2019GT} were adopted for scalable inference. All the aforementioned works design group tests based on the entropy-based utility function. Recently, other utility functions, including mutual information~(MI) and expected area under the receiver operating characteristic curve~(AUC), have been explored with a sequential Monte-Carlo (SMC) method in~\cite{Google2020GT}.  Capacity for noiseless group testing was given in~\cite{Baldassini2013GT}. 
To the best of our knowledge, theoretical analysis for noisy adaptive group testing is limited. In~\cite{Bai2019GT}, the sample complexity when using the entropy utility function was derived. However, computational of the sample complexity requires the ground-truth probability, which is typically unknown in practice. 

In this paper, we consider adaptive group testing in the presence of uncertainty. We follow the model formulation in~\cite{Google2020GT} to set a Bernoulli prior for the infection status. Group testing design is based on a mutual information utility function of the current posterior, iteratively updated given previous results. Based on a stopping criterion of conditional entropy, we derive a lower bound $T_E$ of the required number of group tests. More importantly, we further analyze the sample complexity when we have possibly mismatched testing models. We prove that when the model parameters are mismatched with the ground truth, 
the lower bound increases as expected, $T_E'=(1+\alpha) T_E$ with a constant $\alpha>0$, due to the optimal group selection based on the biased utility function. 
Such a theoretical analysis has not been discussed in the existing literature. We further confirm our analyses by simulation results.

\section{Problem Setup}
\label{sec:format}

Given a population of $n$ individuals whose infection statuses are unknown and modeled by Bernoulli random variables $X_i$'s, $X_i = 1$ if the $i$-th individual is infected, and otherwise $X_i = 0$. 
Denote the random vector representing our understanding on the population infection state by $\textbf{X} = (X_1, X_2, \dots, X_n),$ where $X_i \sim \mbox{Bern}(q_i)$ with the infection probability $q_i$ for the $i$th individual. We are interested in designing group tests adaptively to discover the unknown infection state: $\mathbf{x_{true}} \in \{0, 1\}^n$. 
We assume that the Bernoulli random variables $X_i$'s are independent, i.e. for any $\textbf{x} =(x_1, x_2, \ldots, x_n)\in \{0, 1\}^n$, we can denote $\pi_0(\textbf{x}) = P(\textbf{X} = \textbf{x}) = \prod_i q_i^{x_i} (1-q_i)^{1-x_i}$. Here, $P(X_i=1) = q_i$ can be identical or vary across individuals. 

Throughout the paper, we will use capital letters to denote random variables and vectors are with the bold font.

\subsection{Group Testing}


For time and cost efficiency, instead of testing each individual, we test gathered samples mixed from the selected individuals (saliva samples, for example) to discover the infection status. For the $j$-th group test result $Y_j \in \{0, 1\}$, $Y_j = 1$ if the $j$-th sample is tested positive, indicating that the combined sample contains the sample(s) from infected individual(s); and otherwise all the individuals drawn in the $j$th group test are not infected. 
Group testing design is to choose a subset of individuals from the population as a group, denoted as a vector $\textbf{g}_j = (g_{j ,1}, g_{j ,2}, \ldots, g_{j ,n}) \in \{0, 1\}^n$, and test the mixed samples from the individuals with the corresponding $g_{j, i} = 1$. 


\subsection{Group Testing Parameters}

Existing testing assays have limitations and it is possible to have testing errors. As in~\cite{Google2020GT}, we assume that the group testing has the following sensitivity ($s_{\textbf{g}_j}$) and specificity ($\sigma_{\textbf{g}_j}$): 
\begin{equation}\label{eq:s_def}
    P(Y_j = 1|[\textbf{g}_j, \textbf{x}] = 1) = s_{\textbf{g}_j},
\end{equation}
\begin{equation}\label{eq:sig_def}
    P(Y_j = 0|[\textbf{g}_j, \textbf{x}] = 0) = \sigma_{\textbf{g}_j}
\end{equation}
where $[\textbf{g}, \textbf{x}] = \min(1, \textbf{g}^T\textbf{x}) \in \{0, 1\}$. 
Here, $s_{\textbf{g}_j}$ and $\sigma_{\textbf{g}_j}$ are referred as model parameters for adaptive group testing design in the following part of the paper. 


\subsection{Inference}


Assume that we have designed a batch of testing groups $\mathcal{G}_t = \{\textbf{g}_{t_m + 1}, \dots, \textbf{g}_{t_m + m}\}$ at stage $t$, where $t_m = (t-1)m$ and $m$ is the batch size. Given their corresponding test results $\textbf{Y}_t = \{Y_{t_m + 1}, \dots, Y_{t_m + m}\}$, we can compute: 
\begin{equation}\label{eq:y_given_x}
    Pr(\textbf{Y}_t = \textbf{y}_t|\textbf{X} = \textbf{x}) = \prod_{j=1}^m (Q_{t_m + j}^{(0)})^{(1-y_t^{(j)})} (Q_{t_m + j}^{(1)})^{y_t^{(j)}} 
\end{equation}
based on~\eqref{eq:s_def} and~\eqref{eq:sig_def}, where 
\begin{eqnarray}
    && Q_{t_m + j}^{(0)} = \sigma_{\textbf{g}_{t_m + j}} - \rho_{\textbf{g}_{t_m + j}}[\textbf{g}_{t_m + j}, \textbf{x}], \\
\mbox{and  } &&    Q_{t_m + j}^{(1)} = 1-\sigma_{\textbf{g}_{t_m + j}} + \rho_{\textbf{g}_{t_m + j}}[\textbf{g}_{t_m + j}, \textbf{x}]. \end{eqnarray}
Here, $\rho_{\textbf{g}_{t_m + j}} = s_{\textbf{g}_{t_m + j}} + \sigma_{\textbf{g}_{t_m + j}} -1$.

We can further infer the posterior of the population infection status by Bayes's rule: 
\begin{align}\label{eq:pi_adaptive}
\begin{split}
    \pi_t(\textbf{x}) &= Pr(\textbf{X} = \textbf{x}|\textbf{Y}_{1} = \textbf{y}_1, \textbf{Y}_{2} = \textbf{y}_2, \dots, \textbf{Y}_{t} = \textbf{y}_t)\\
    &\propto Pr(\textbf{X} = \textbf{x}, \textbf{Y}_{1} = \textbf{y}_1, \textbf{Y}_{2} = \textbf{y}_2, \dots, \textbf{Y}_{t} = \textbf{y}_t)\\
    &= \pi_0(\textbf{x}) \prod_{k=1}^t Pr(\textbf{Y}_{k} = \textbf{y}_k|\textbf{X} = \textbf{x}).
\end{split}
\end{align}

For simplicity, from now on, we write the posterior of an event $E$ given the previous $t$ test results by: %
\begin{equation}\label{eq:pt_def}
    P_{t-1}(E) = P(E|\textbf{Y}_{1} = \textbf{y}_1, \textbf{Y}_{2} = \textbf{y}_2, \dots, \textbf{Y}_{t-1} = \textbf{y}_{t-1}).
\end{equation}

\section{Adaptive Group Testing}

For adaptive group testing, we actively select a batch of groups to update the posterior $\pi_t$, and design a utility function $U_t(\mathcal{G}_t) = U(\mathcal{G}_t, \pi_t)$ to guide the group selection in the next iteration.
More specifically, the task at each stage is to find a batch of groups $\mathcal{G}_t^*$ such that 
\begin{equation}
    \mathcal{G}_t^* \in \mathop{\arg\max}_{\mathcal{G}_t} U_t(\mathcal{G}_t). 
\end{equation}

\subsection{Mutual Information Utility}

One of natural choices of the utility function is the Mutual Information: 
\begin{equation}\label{eq:prob}
    U_{MI}(\mathcal{G}_t, \pi_t) = I(\textbf{X}; \textbf{Y}_{k}| \textbf{Y}_{1}, \dots, \textbf{Y}_{k-1}). 
\end{equation}

Denote $h(p) = - p \log_2 p - (1 - p) \log_2(1 - p)$ as the binary entropy and denote $I(\textbf{X}; \textbf{Y}_{k}| \textbf{Y}_{1}, \textbf{Y}_{2}, \dots, \textbf{Y}_{k-1})$ as $I_{P_{k-1}}(\textbf{X}; \textbf{Y}_{k})$. Write $H(\textbf{X}| Y_{1}, Y_{2}, \dots, Y_{{t}})$ as $H_{P_t}(\textbf{X})$. According to~\eqref{eq:pi_adaptive} the Mutual Information can be written as 
\begin{equation}\label{eq:ce}
\begin{split}
    &I_{P_{t-1}}(\textbf{X};\textbf{Y}_t) = H_{P_{t-1}}(\textbf{Y}_t) \\
    &\quad\quad - \sum_{j=1}^k{[h_{\sigma_{\textbf{g}_{t_m + j}}} + \gamma_{\textbf{g}_{t_m + j}}f_{\pi_t}(\textbf{g}_{t_m + j})]},
    \end{split}
\end{equation}
where $h_{\phi} = h(\phi)$,  $\gamma_{\textbf{g}} = h_{s_{\textbf{g}}} - h_{\sigma_{\textbf{g}}}$, and
\begin{equation}\label{eq:f}
f_{\pi_t}(\textbf{g}) = \sum_\textbf{x}{\pi_t(\textbf{x})[\textbf{g}, \textbf{x}]} = \sum_{\textbf{x}: [\textbf{g}, \textbf{x}]=1}{\pi_t(\textbf{x})}
\end{equation}
represents the probability of having infected patient(s) in the chosen group.

Here, we consider the simplified setups with all the sensitivity and specificity being constant with respect to group selection, i.e. $\sigma_{\textbf{g}_{t_m + j}} = \sigma$, and $s_{\textbf{g}_{t_m + j}} = s$. 
The Mutual Information \eqref{eq:ce} can be written as: 
\begin{equation}\label{eq:ce2}
    I_{P_{t-1}}(\textbf{X};\textbf{Y}_t) = H_{P_{t-1}}(\textbf{Y}_t) - \sum_{j=1}^k{[h_{\sigma} + \gamma f_{\pi_t}(\textbf{g}_{t_m + j})]}. \end{equation}
Further assume that we test one group at each stage. We have: 
\begin{equation}\label{eq:ce_one_group}
    I_{P_{t-1}}(\textbf{X};Y_t) = h(\rho f_{\pi_t}(\textbf{g}_t) + 1 - \sigma) - h_{\sigma} - \gamma f_{\pi_t}(\textbf{g}_t), 
\end{equation}
where $ \rho f_{\pi_t}(\textbf{g}) + 1 - \sigma = P_{t}(Y_t = 1)$. Note that we have replaced $\textbf{Y}_t$ by $Y_t$ and $\mathcal{G}_t $ by $\textbf{g}_t$. Write $J(x) = h(\rho x + 1 - \sigma) - h_{\sigma} - \gamma x$, \eqref{eq:ce_one_group} can be written as: 
\begin{equation}\label{eq:ce_one_group_2}
    I_{P_{t-1}}(\textbf{X};Y_t) = J(f_{\pi_t}(\textbf{g}_t)),
\end{equation}
where $J(x)$ is a concave function of $x$ so it would be maximized if its derivative is zero, leading to the closed-form optimal point of $f_{\pi_t}(\textbf{g}_t)$ for~\eqref{eq:ce_one_group}:
\begin{equation}\label{eq:f_star}
    f^* = \frac{\sigma}{\rho} - \frac{\exp{\frac{\gamma}{\rho}}}{\rho (\exp{\frac{\gamma}{\rho}} + 1)}. 
\end{equation}

Note that $J(x)$ is fixed when the group testing sensitivity and specificity parameters are given. As it is concave, it is easy to find $f_{\pi_t}(\textbf{g}_t)$ that optimizes $J(x)$.

The problem defined in~\eqref{eq:prob} becomes
\begin{equation}
    \textbf{g}_t^* \in \mathop{\arg\max}_{\textbf{g}_t} U_t(\textbf{g}_t), 
\end{equation}
and can be informally viewed as finding $\textbf{g}_t^*$ to make $f_{\pi_t}(\textbf{g}_t^*)$ as close as possible to $f^*$.


    
    
\subsection{Stopping Criteria}
In previous works, either the budget \cite{Google2020GT} or the maximum probability of infection status, i.e. $\max_\textbf{x}{P_{t-1}(\textbf{x})}$ \cite{Bai2019GT, Sakata2021GT}, have been used as stopping criteria. The former cannot help us analyze the asymptotic performance, while the latter can be tricky to analyze if using estimation methods like LBP. So here we use Conditional Entropy (CE) as the stopping criterion: 
\begin{equation}\label{eq:stpo}
    H(\textbf{X}| Y_{1}, Y_{2}, \dots, Y_{{t}}) \leq \delta H(\textbf{X}), 
\end{equation} 
where $0\leq\delta\leq 1$ and $H(\textbf{X})$ is the entropy of the prior and is fixed once the prior is given. 

With the definition of mutual information and~\eqref{eq:pi_adaptive}, we have
\begin{align}\label{eq:entropy_sum}
\begin{split}
    & I(\textbf{X}; Y_{1}, Y_{2}, \dots, Y_{T}) \\
    =& H(\textbf{X}) - H(\textbf{X}| Y_{1}, Y_{2}, \dots, Y_{T})\\
    =& \sum_{k=1}^t[H(\textbf{X}| Y_{1}, Y_{2}, \dots, Y_{{k-1}}) - H(\textbf{X}| Y_{1}, Y_{2}, \dots, Y_{{k}})]\\
    =& \sum_{k=1}^t I_{P_{k-1}}(\textbf{X}; Y_{{k}}). 
\end{split}
\end{align}
So we have
\begin{equation}\label{eq:entropy_cond_sum}
H_{P_t}(\textbf{X}) = H(\textbf{X}) - \sum_{k=1}^t I_{P_{k-1}}(\textbf{X}; Y_{{k}}). 
\end{equation}
If we only search for one group to query each stage, by substituting~\eqref{eq:ce_one_group} and~\eqref{eq:ce_one_group_2} into~\eqref{eq:entropy_cond_sum},
\begin{equation}\label{eq:entropy_cond_sum_single}
H_{P_t}(\textbf{X}) = H(\textbf{X}) - \sum_{k=1}^t J(f_{\pi_{k}}(\textbf{g}_k)). 
\end{equation}

The stopping criterion becomes 
\begin{equation}\label{eq:stp}
    \sum_{k=1}^t J(f_{\pi_{k}}(\textbf{g}_k)) \geq (1 - \delta) H(\textbf{X}). 
\end{equation}

By doing this, we are able to give the stopping criterion informational meaning and it is easy to compute once each $f_{\pi_t}(\textbf{g}_t)$ is given by group selection.

\subsection{Sample Complexity}
It worth noting that by the nature of deriving $f(\textbf{g})$ in~\eqref{eq:f}, we are not guaranteed to reach $f^*$ every time during adaptive group testing. In other words, there are always gaps between the achieved $f(\textbf{g})$ by group test design and the optimal $f^*$. Therefore we cannot treat the information gain at each iteration as a constant. Besides, it can be difficult to analyze how close $f_{\pi_t}(\textbf{g})$ can be approach $f^*$ as $\pi_t$ adapts over the iterations.

\begin{figure}[t]
    \centering
    \includegraphics[width=0.4\textwidth]{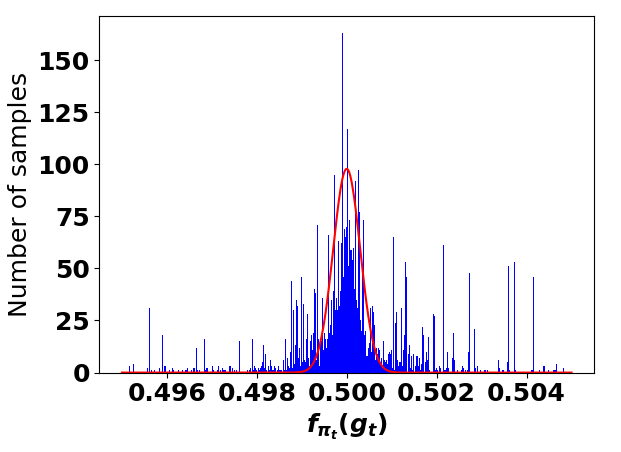}
    \caption{The histogram of selected $f_{\pi_t}(\textbf{g}_t)$ in the 5-th to 15-th iteration in the experiment. The red line is the PDF of $\mathcal{N}(0.5, (3e-4)^2)$. We can see that the distribution of $f_{\pi_t}(\textbf{g}_t)$ can be reasonably fitted by a Gaussian distribution. The group selection of the first few iterations would be highly influenced by the prior so we do not include them here.}
    \label{fig:f_gauss}
\end{figure}

Here, we approximate $f_{\pi_t}(\textbf{g}_t^*) \in \mathop{\arg\max}_{f_{\pi_t}(\textbf{g}_t)} J(f_{\pi_t}(\textbf{g}_t))$ as normal distributions $F_t \sim N(f^*, \nu_{F}^2)$. An example histogram of the selected $f_{\pi_t}(\textbf{g}_t)$ in the experiments is illustrated in Figure \ref{fig:f_gauss}.

Thus, we can transform the stopping criterion~\eqref{eq:stp} into 
\begin{equation}
    \sum_{k=1}^t J(F_t) \geq (1-\delta)H(\textbf{X}). 
\end{equation}

Note that 
\begin{equation}
    h(x) \geq - 4(x - 0.5)^2 + 1.
\end{equation}


Therefore,  
\begin{equation}\label{eq:two_sides}
    J(x) \geq J^{(4)}(x),
\end{equation} 
where $J^{(A)}(x) = -A\rho^2x^2 - [2A(0.5-\sigma)\rho - \gamma]x - A(0.5 - \sigma)^2 + 1 - h_\sigma$. 

\begin{theorem}\label{th:1}
If $T \geq T_E^{(A)}$,
we have 
\begin{align}
\begin{split}
    &Pr(\sum_{k=1}^T J^{(A)}(F) \geq (1-\delta)H(\textbf{X})) \\ &\quad\quad \geq \frac{[T E_F^{(A)} - (1-\delta)H(\textbf{X})]^2}{T V_F^{(A)} + [T E_F^{(A)} - (1-\delta)H(\textbf{X})]^2}, 
\end{split}
\end{align}
where $T_E^{(A)} = \frac{(1-\delta)H(\textbf{X})}{E_F^{(A)}}$, $E_F^{(A)} = -A\rho^2 \nu_{F}^2 - A(0.5 - \sigma)^2 + 1 - h_\sigma -A\rho^2 (f^*)^2 - B_Af^*$, $V_F^{(A)} = 2A^2\rho^4\nu_{F}^4 + (B_A + 2A\rho^2 f^*)^2\nu_{F}^2$, and $B_A = 2A(0.5-\sigma)\rho - \gamma$.
\end{theorem}

\begin{proof}
When $F \sim N(f^*, \nu_{F}^2)$, 
\begin{align}
\begin{split}
    &J^{(A)}(F)
    =-A\rho^2 N^2 - (B_A + 2A\rho^2f^*)N  \\
    &\quad - A(0.5 - \sigma)^2 + 1 - h_\sigma -A\rho^2 (f^*)^2 - B_Af^*,
\end{split}
\end{align}
where $N \sim N(0, \nu_{F}^2)$ and $B_A = 2A(0.5-\sigma)\rho - \gamma$. 

Denote 
\begin{equation}
\begin{split}
    E_F^{(A)}
    =& E[J^{(A)}(F)] \\
    =& -A\rho^2 \nu_{F}^2 - A(0.5 - \sigma)^2 + 1\\
    &\quad\quad- h_\sigma -A\rho^2 (f^*)^2 - B_Af^*,
\end{split}
\end{equation} 
and 
\begin{equation}
\begin{split}
    V_F^{(A)} =& Var[J^{(A)}(F)]\\
    =& 2A^2\rho^4\nu_{F}^4 + (B_A + 2A\rho^2 f^*)^2\nu_{F}^2. 
\end{split}
\end{equation}

Considering $F_t$ as independent sampled, we also have $J(F_t)$ are independent. So that we have 
\begin{equation}
     E[\sum_{k=1}^T J^{(A)}(F_k)] = T E_F^{(A)}
\end{equation} 
and 
\begin{equation}
    Var[\sum_{k=1}^T J^{(A)}(F_k)] = T V_F^{(A)}.
\end{equation}

By Chebyshev's inequality,
if $T E_F^{(A)} \geq (1-\delta)H(\textbf{X})$, or $T \geq T^{(A)} = \frac{(1-\delta)H(\textbf{X})}{E_F^{(A)}}$, 
\begin{align}
\begin{split}
    & Pr(\sum_{k=1}^T J^{(A)}(F) \geq (1-\delta)H(\textbf{X})) \\
    & \quad\quad  \geq \frac{[T E_F^{(A)} - (1-\delta)H(\textbf{X})]^2}{T V_F^{(A)} + [T E_F^{(A)} - (1-\delta)H(\textbf{X})]^2}.
\end{split}
\end{align}
\end{proof}

We now give the condition for $H_{P_T}(\textbf{X}) \leq \delta H(\textbf{X})$ 
 with the following lemma.

\begin{prop}\label{lm:1}
If $T \geq T_E^{(4)}$, we have 
\begin{align}
    \begin{split}
        &Pr(H_{P_T}(\textbf{X}) \leq \delta H(\textbf{X})) \\
        &\quad\quad \geq 1 - \frac{T V_F^{(4)}}{T V_F^{(4)} + [(T - T^{(4)}) E_F^{(4)}]^2}.
    \end{split}
\end{align}
\end{prop}

\begin{proof}
From Equation \ref{eq:two_sides} and Theorem \ref{th:1}, we have this lemma.
\end{proof}

Based on this theorem, we have shown that the probability of meeting the stopping criterion is in the rate of $1 - o(T^{-1})$ when $T \geq T_E^{(4)}$. $T_E^{(4)}$ takes the form $\frac{(1-\delta)H(\textbf{X})}{E_F^{(A)}}$, which is related to the prior distribution. When setting i.i.d. Bernoulli prior, $T_E^{(4)}$ becomes $n\frac{(1-\delta)h(q)}{E_F^{(A)}}$, which is proportional to the number of patients. In general the variance $\nu_{F} \ll 1$ is small, which leads to $(E_F^{(4)})^2 \gg V_F^{(4)}$ so that $Pr(H_{P_T}(\textbf{X}) \leq \delta H(\textbf{X}))$ can be close to $1$ very quickly as soon as $T \geq T_E^{(4)}$.

\section{Mismatched Model}
\label{sec:pagestyle}
Now consider we have biased assumptions on the group test model parameters. Specifically, the true sensitivity and specificity is $s$ and $\sigma$, but we do not have them in practice and set them as $s'$ and $\sigma'$ (mismatched) for adaptive group testing. 

In each iteration, we would optimize the `mismatched' utility function: 
\begin{align}\label{eq:ce_mis}
\begin{split}
    &I_{P_{t-1}'}'(\textbf{X};\textbf{Y}_t) = H_{P_{t-1}'}(\textbf{Y}_t) \\
    & \quad\quad- \sum_{j=1}^k{[h_{\sigma_{\textbf{g}_{t_m + j}}}' + \gamma_{\textbf{g}_{t_m + j}}'f_{\pi_t'}(\textbf{g}_{t_m + j})]}
\end{split}
\end{align}
and select the group such that $\mathcal{G}' \in \mathop{\arg\max}_\mathcal{G} I_{P_{t-1}'}'(\textbf{X};\textbf{Y}_t)$, 
where $\pi_t'$ is the mismatched posterior updated with the mismatched parameters. With the same setup in Section 3, the selection at each iteration is
$\textbf{g}' \in \mathop{\arg\max}_\textbf{g} J'(f_{\pi_{t}'}(\textbf{g}))$. 


The mismatched selection target of $f_{\pi_t'}(\textbf{g})$ would be
\begin{equation}
    f' = \frac{\sigma'}{\rho'} - \frac{\exp{\frac{\gamma'}{\rho'}}}{\rho' (\exp{\frac{\gamma'}{\rho'}} + 1)}.
\end{equation}

\begin{figure}[t]
    \centering
    \includegraphics[width=0.5\linewidth]{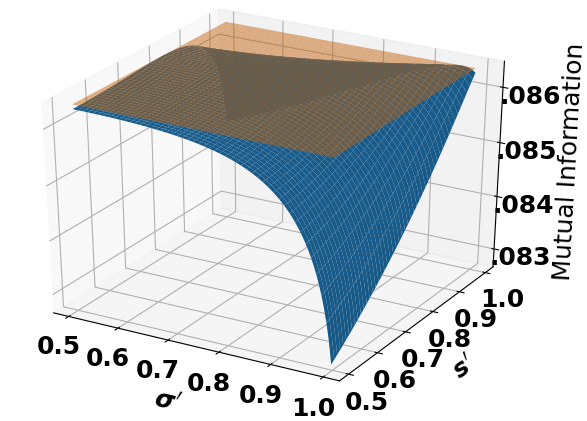}
    \caption{MI utility with the change of $\sigma'$ and $s'$ when we set ground truth $\sigma=0.6$, $s=0.8$. The z axis is $J(f')$. The orange plane represents the optimal Information Gain. }
    \label{fig:MI}
\end{figure}

The actual information gain, however, should be calculated with the true parameters, 
\begin{equation}
    I_{P_{t-1}}(\textbf{X};\textbf{Y}_t) = J(f_{\pi_t}(\textbf{g}_t')). 
\end{equation}

Notice that here the true posterior $\pi_t$ needs to be updated with true parameters and have the `true' understanding on the infection status.

\begin{figure}[t]
 \centering
    \begin{subfigure}{0.2\textwidth}
    \includegraphics[width=\linewidth]{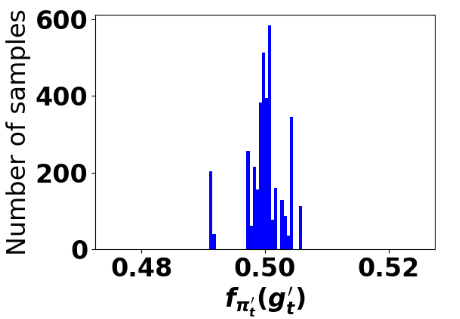}
    \subcaption{Histogram of $f_{\pi_t'}(\textbf{g}_t')$}
    \label{subfig:mis1}
    \end{subfigure}
    \begin{subfigure}{0.2\textwidth}
    \includegraphics[width=\linewidth]{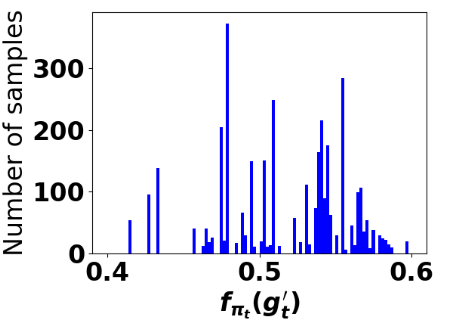}
    \subcaption{Histogram of $f_{\pi_t}(\textbf{g}_t')$}
    \label{subfig:mis2}
    \end{subfigure}
    \caption{The histogram of selected $f_{\pi_t}(\textbf{g}_t)$ in the 3-th to 7-th iteration of experiment  over 1000 runs.}
\end{figure}

Similar as what we derive in the previous section, we consider $f_{\pi_t}'(\textbf{g}_t) = F_t' \sim N(f', (\nu_{F'})^2)$. A histogram in a example is illustrated in Figures~\ref{subfig:mis1} and~\ref{subfig:mis2}. When the model is mismatched, the variance would be much larger. 

Similar to Lemma \ref{lm:1}, we have: 

\begin{prop}
If $T \geq (1+\alpha)T^{(4)}$, we have 
\begin{align}
\begin{split}
    &Pr(H_{P_T}(\textbf{X}) \leq \delta H(\textbf{X})) \\
    &\quad\quad \geq 1 - \frac{T (V_F^{(4)})'}{T (V_F^{(4)})' + [(T - (1+\alpha^{(4)})T^{(4)}) (E_F^{(4)})']^2}
\end{split}
\end{align}
for mismatched models, where 

$\alpha^{(A)} = \frac{A\rho^2 (f')^2 + B_Af' -A\rho^2 (f^*)^2 - B_Af^*}{E_{F'}^{(A)}}$, 

$E_{F'}^{(A)} = -A\rho^2 (\nu_{F'})^2 - A(0.5 - \sigma)^2 + 1 - h_\sigma -A\rho^2 (f')^2 - B_Af'$, $(V_F^{(A)})' = 2A^2\rho^4(\nu_{F'})^4 + (B_A + 2A\rho^2 f')^2(\nu_{F'})^2$. 
\end{prop}

\begin{proof}
The proof can be easily adopted from the proof of Lemma \ref{lm:1}, simply by replacing $F_t$ with $F_t'$.
\end{proof}

Here $\alpha^{(4)}$ represents the change of sample complexity because of bias on $f'$ to $f^*$. We want to point out that $((E^{(4)})')^2 \gg \nu_{F'} \gg \nu$ holds, so the probability can still be close to 1 once $T \geq (1+\alpha)T^{(4)}_E$. The main influence of biased model is the difference of $f'$ and $f^*$. Also $\alpha^{(4)} = 0$ if $f' = f^*$, so that we can observe that the performance is similar when $f' = f^*$ in the experiments.

\section{Experimental Results}
\label{sec:typestyle}

We perform simulations to confirm the derived bounds of the required number of group testing iterations in this section. 

\subsection{Experimental Settings}

We investigate  how the group testing performance changes with different model parameter settings. We have simulated results for 24 combinations of mismatched parameters (biased) $\sigma', s' \in \{0.6, 0.7, 0.8, 0.9, 0.99\}$ together with the ground-truth group testing parameters (unbiased), $\sigma' = \sigma = 0.8$ and $s' = s = 0.8$. To allow the exhaustive search to achieve the best achieved group design, we have simulated 1000 runs with a population of ten individuals with one infected individual. The prior $\pi_0$ is set as the independent Bernoulli for each individual and the probability of each individual being infected is $0.1$, $X_i \sim \mbox{Bern}(0.1)$. We perform adaptive group test as described in previous sections for each simulation run and take the average conditional entropy and Area Under ROC Curve (AUC)~\cite{Bradley1997AUC} over 1000 runs for each iteration for performance evaluation. 

\subsection{Conditional Entropy}

\begin{figure}
\centering
\begin{subfigure}{0.2\textwidth}
\includegraphics[width=\linewidth]{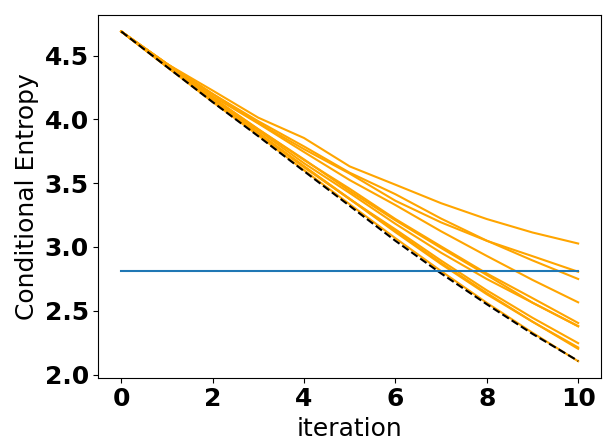}
\caption{Average CE}
\label{subfig:CE}
\end{subfigure}
\begin{subfigure}{0.2\textwidth}
\includegraphics[width=\linewidth]{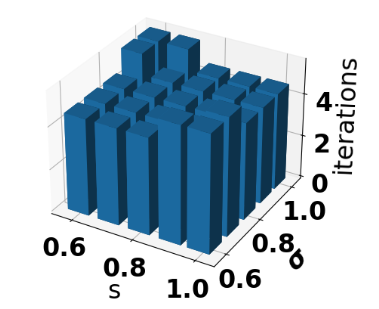}
\caption{$\delta=0.8$}
\label{subfig:8}
\end{subfigure}
\begin{subfigure}{0.2\textwidth}
\includegraphics[width=\linewidth]{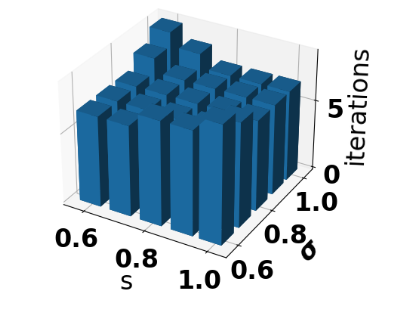}
\caption{$\delta=0.7$}
\label{subfig:7}
\end{subfigure}
\begin{subfigure}{0.2\textwidth}
\includegraphics[width=\linewidth]{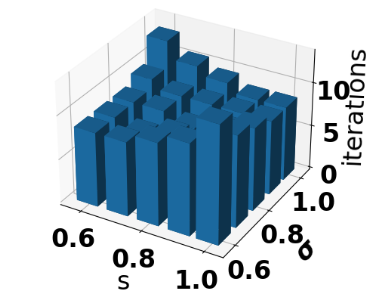}
\caption{$\delta=0.6$}
\label{subfig:6}
\end{subfigure}
\caption{Simulation Results}
\end{figure}

In this set of experiments, the ground-truth entropy for performed simulations is: 
$$H(\textbf{X}) = n [p\log{p}+(1-p)\log{(1-p)}] = \log_2{0.1} + 9\log_2{0.9}. $$
We have plotted the average Conditional Entropy defined in~\eqref{eq:entropy_cond_sum_single} over iterations in Figure \ref{subfig:CE}.
The dashed curve is the performance based on the ground-truth model parameters, which outperforms the ones based on the utility function with mismatched models as we expected.


Then we show the average of the required group test number in Figures~\ref{subfig:8}, \ref{subfig:7}, and \ref{subfig:6} when $\delta = 0.8, 0.7, 0.6$, respectively. The three horizontal lines in Figure \ref{subfig:CE} shows the value of $\delta H(\textbf{X})$ when $\delta = 0.8, 0.7, 0.6$. We can see that when $f' = f^*$, we have similar required test numbers.

\subsection{AUC}

Here, we compute the area-under-curve (AUC) based on the marginal likelihood as the criterion to evaluate the performance of our updated posterior given corresponding group test results in each iteration. 

We compute the marginal likelihood for each of the individuals, indexed by $i$, as 
$$m_t(i) = P_{t-1}(X_i = 1) = \sum_{[\textbf{g}_i, \textbf{x}] = 1}\pi_t(\textbf{x}),$$
where $X_i$ is the infection status of the $i$-th individual, $\textbf{g}_i$ is a group that only contains $i$-th individual, i.e. one-hot coding of the $i$-th individual. 

\begin{figure}[t]
\centering
\begin{subfigure}{0.2\textwidth}
\includegraphics[width=\linewidth]{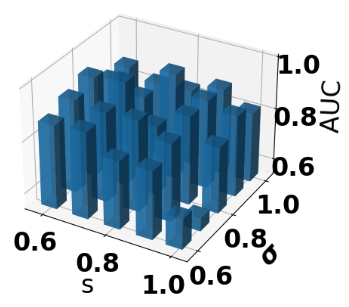}
\caption{AUC after 4 tests}
\label{subfig:AUC4}
\end{subfigure}
\begin{subfigure}{0.2\textwidth}
\includegraphics[width=\linewidth]{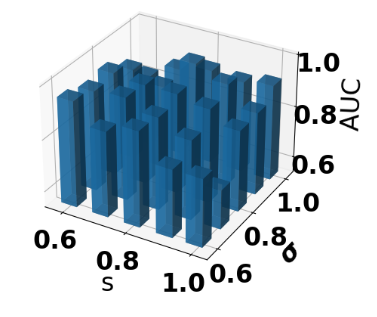}
\caption{AUC after 8 tests}
\label{subfig:AUC8}
\end{subfigure}
\caption{Average AUC after different numbers of tests in different setups}
\end{figure}

The AUC metric of a classifier $f$ is defined as 
\begin{equation}
    \textbf{AUC}(f) = E_{x^+ \in POS} E_{x^- \in NEG} \mathbb{I}(f(x^+)>f(x^-)).
\end{equation}
The AUC of the infection marginal likelihood $m_t$ can be written as 
\begin{equation}
\textbf{AUC}(m_t) = \frac{\sum_{i^+ \in POS} \sum_{i^- \in NEG}\mathbb{I}(m_t(i^+)>m_t(i^-)) }{|POS||NEG|},
\end{equation}
where $\mathbb{I}$ is the indicator function: $\mathbb{I}(E) = 1$ if $A$ is true, and $\mathbb{I}(E) = 0$ otherwise.

Although not directly optimized over AUC, the AUC values for each setup in given iterations illustrated in Figures \ref{subfig:AUC4} and \ref{subfig:AUC8} show that conditional entropy is relevant, though not strictly monotonically, to the accuracy of infection detection.

\section{Conclusions}
In this paper, we proved that the probability of meeting the stopping criterion based on conditional entropy is in the rate of $1-o(T^{-1})$. More importantly, we have shown that a mismatch in the group testing model would lead to a multiplicative constant $1+\alpha$ ($\alpha > 0$), determined by the difference between $f^*$ and $f'$, to the required number of group tests. 
Our simulation study shows that the adaptive group testing can be efficient in infection detection based on the mutual information utility. Adaptive design with the correct group testing model outperforms the ones with mismatched models. 
The performance evaluation by AUC has shown to be related to the conditional entropy though not strictly monotonic. 

\vfill\pagebreak

\bibliographystyle{IEEEbib}
\bibliography{citation}

\end{document}